\documentclass[prc,aps,twocolumn,amssymb,aps,nofootinbib,floatfix]{revtex4}
\usepackage[ansinew]{inputenc}
\usepackage[T1]{fontenc} 
\usepackage{graphicx}
\usepackage{epsfig}

\newcommand{\MeV}{\hbox{ MeV}}
\newcommand{\GeV}{\hbox{ GeV}}
\newcommand{\TeV}{\hbox{ TeV}}
\newcommand{\fm}{\hbox{ fm}} 
\newcommand{\atanh}{\hbox{ atanh}} 

\begin{document}

\title{Decreasing elliptic flow at the LHC: Calculations in a parton recombination approach}

\author{Daniel Krieg, Marcus Bleicher}
\affiliation{Institut f\"ur Theoretische Physik, Johann Wolfgang Goethe-Universit\"at, 
Max-von-Laue-Str.~1, 60438 Frankfurt am Main, Germany}
\date{September 4, 2007}

\begin{abstract}
We apply the parton recombination approach to study the energy dependence of the elliptic flow, $v_2$ in heavy ion 
collisions from AGS to LHC energies. The relevant input quantities ($T, \mu_B, \eta_T$) at the various center of 
mass energies are obtained from fits to the available data. The model yields a good description of the integrated
$v_2$ data for charged particles at midrapidity from AGS to RHIC energies. In stark contrast to the current expectations,
we observe a decrease of the integrated $v_2$ values above the highest RHIC energy. Thus, we predict a decrease of $v_2$
at LHC energies compared to the RHIC results. This drop is attributed to negative $v_2$ values for the underlying 
parton distributions at low to moderate transverse momenta that develops if the transverse flow velocity is high enough. 
At energies above the LHC regime, the present approach predicts even negative values for the integrated $v_2$.
\end{abstract}

\maketitle

The main goal of the current and past heavy ion programs is the search for a new state of matter
called the Quark-Gluon-Plasma (QGP) \cite{Bass:1998vz}. Major breakthroughs  for the 
potential discovery \cite{Adcox:2004mh,Adams:2005dq}
of this new state of matter where (I) the experimental discovery of a large elliptic flow ($v_2$) at RHIC. 
This lead to the conclusion, that the matter formed in the early stage of a heavy ion reaction at 
RHIC behaves like a nearly perfect liquid (i.e. a liquid with a low viscosity) (\cite{Kolb:2003dz} and Refs. therein). 
(II) The observation of constituent
quark number scaling $v_2^{\rm hadron} (p_T^{\rm hadron}) = n_q v_2^{q} (p_T^{\rm hadron}/n_q)$, meaning that the 
elliptic flow of baryons (three quarks, $n_q=3$) versus mesons (two quarks, $n_q=2$)) 
scales like $3:2$ at the hadron transverse momentum $p_T^{\rm hadron}$. While the details of this scaling are 
still under discussion \cite{Greco:2004yc,Lin:2004en,Molnar:2005wf,Lu:2006qn,Hwa:2007ae}, one might generally see this 
scaling as evidence for a recombination like hadronization process for the transition of partonic matter 
to hadronic matter \cite{Zimanyi:1999py,Fries:2003kq}.

In this letter, we will apply the recombination approach to explore the energy dependence of the elliptic flow
in massive (Pb+Pb/Au+Au) nucleus-nucleus reactions from the AGS energies to the highest LHC energy. 
For complemantory explorations of the elliptic flow at LHC within transport approaches the reader is referred to
\cite{Ko:2007zzc,Molnar:2007an,Xu:2007jv}. 
Elliptic flow is a well chosen observable for the exploration of the energy dependence of flow observables, because 
it exhibits a self quenching effect and is therefore mostly sensitive to the early (partonic) stage of the reaction, even
at rather low beam energies. The structure of this letter is as follows: We start with a summary of the
energy dependence of the input parameters, then we discuss shortly the relevant recombination formulas, finally we present the results for the elliptic flow excitation function and the predictions for LHC.

To apply this model to other energies than RHIC ($\sqrt{s} = 200 \GeV$) one has to model the dependence 
of the temperature $T$, the baryo-chemical potential $\mu_B$ and the transverse flow 
rapidity $\eta_T=\atanh(\beta_T)$ at the hadronization surface as a function of the center of mass energy. 
Details of the parameterisations can be found in the appendix. The values for RHIC energies and below are 
fitted to previously extracted flow velocities, at LHC energies, we obtain the values 
(for $\sqrt{s_{NN}}=5.5 \TeV$): $T=175 \MeV$, $\mu_B \simeq = 0 \MeV$ and $\beta_T=0.75 c$, 
which are in line with previous estimates \cite{Fries:2003fr}.

With these parameters we predict the elliptic flow of hadrons using the collinear recombination 
approach \cite{Fries:2003kq}. Since we are presently only interested in bulk (i.e. low $p_T$ quantities, 
we apply the recombination prescription in the low to intermediate $p_T$-range from 0 to 
about 3~GeV at RHIC and about 5-6~GeV at LHC respectively, above this momentum, fragmentation will 
dominate the underlying quark distribution. Comparing the yields to experiment the formalism seems 
applicable down to $p_T\sim 1 \GeV$ at RHIC, which is approximately 3\% of the matter produced, and 
we expect this lower limit to rise slightly to $p_T\sim 1.2 \GeV$ at LHC. So using the recombination 
approach down to $p_T\rightarrow 0$ seems questionable, however, a part of the uncertainty in the 
recombination mechanism at low $p_T$, introduced by the violation of energy conservation, cancels 
after taking the ratios in Eq. (\ref{eqn:v2_definition}). Thus, the recombination formalism seems 
to give valid results for $v_2$ down to transverse momenta of several hundred MeV \cite{Fries:2003kq}. 
Further justification about the validity of our results can be found in the paragraph about the mean $v_2$.

The picture that one quark and one anti-quark (or three quarks) with collinear momenta can recombine to form a 
meson (baryon) leads one to an integral over the product of the constituent-quark densities $w_a$ times 
the Wigner function of the hadron $\Phi^W_h$ with the freeze-out hyper-surface $\Sigma$ and its normal 
vector $u^\nu$ \cite{Fries:2003kq}:
\begin{eqnarray}
E \frac{d^3N_h}{dp^3} &=& \int_\Sigma \frac{d\sigma\, P_\nu u^\nu (\sigma)}{(2\pi)^3} \label{eqn:main}\\
&& \times \int \prod_a \frac{d x_a}{(2\pi)^3} \Phi^W_h(x) w_a(R;x_a P) \delta(1-\sum_a x_a)\nonumber
\end{eqnarray}
were the quark $a$ carries the fraction $x_a$ of the hadron momentum $P$.

\begin{figure}
\centering
  \includegraphics{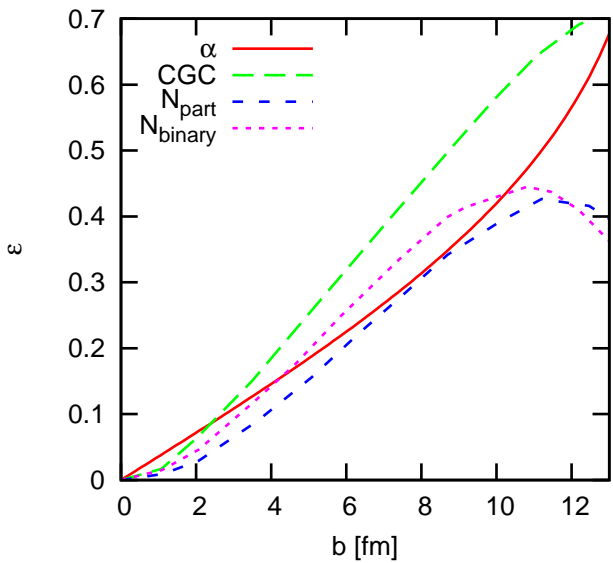}
 \caption{(Color online) Comparison of our asymmetry parameter $\alpha$ to the eccentricity from CGC and the Glauber model \cite{Hirano:2005xf}.}
 \label{plt:asymmetry}
\end{figure}

We use the parameterisations from \cite{Fries:2003kq} with a pure thermal spectrum for the recombining quarks 
and do not include the power-law-tail which is only relevant for fragmentation. At higher cm-energies recombination 
is expected to be dominant up to even higher $p_T$ \cite{Fries:2003fr}, therefore, we neglect the 
contributions from fragmentation for the present considerations. Following \cite{Fries:2003kq} we obtain the elliptic 
flow for non-central reactions from the asymmetry of the overlap region.
To model peripheral collisions the asymmetry $\alpha$ of the collision depends on the geometric width and 
height of the transverse overlap zone
\begin{equation}
h(b)=\sqrt{R_A^2-(b/2)^2}\qquad w(b) =R_A-(b/2)
\end{equation}
as
\begin{equation}
\alpha = \frac{h(b)-w(b)}{h(b)+w(b)}
\end{equation}
We have compared this simple asymmetry parameter with the eccentricity from other 
parameterisations in Fig. (\ref{plt:asymmetry}). One observes that the $\alpha$ parameter used (full line) is 
quantitatively close to the Glauber model (short dashed line) calculation for the relevant impact parameters. 
It has also been speculated that a much 'sharper' Color Glass Condensate initial distribution might be present in the
initial state, this parametrisation is depicted by the long dashed line. The CGC values for the eccentricity are
slightly higher and lead to a small quantitative change of the final $v_2$ values, however the qualitative feature of 
a $v_2$ decrease at high energies is not affected.

This spatial asymmetry is translated into a velocity asymmetry in the parton phase by
\begin{equation}
\eta_T(\phi) = \eta_T^0 \left(1+\alpha f(p_T) \cos(2\phi)\right)
\end{equation}
To account for the fact that faster partons will experience this anisotropy less than slower ones we define
\begin{equation}
f(p_T) = \frac{1}{1+(p_T/p_0)^2}
\end{equation}
with the parameter $p_0=1.1$~GeV  taken from \cite{Fries:2003kq}.
The elliptic flow is then given by
\begin{equation}
v_2 = \left\langle \cos(2 \Phi) \right\rangle 
= \frac{\int \cos(2\Phi) \frac{dN}{dp_t^2} \,d\Phi}{\int \frac{dN}{dp_t^2} \,d\Phi}
\label{eqn:v2_definition}
\end{equation}
and is calculated for the individual quark species as \cite{Fries:2003kq}
\begin{equation}
v_2^q(p_T) = \frac{\int \cos(2\phi) I_2 \left[a(\phi)\right] K_1 \left[b(\phi)\right] \, d\phi}
{\int I_0 \left[a(\phi)\right] K_1\left[b(\phi)\right] \, d\phi}
\end{equation}
\begin{figure}
\centering
  \includegraphics{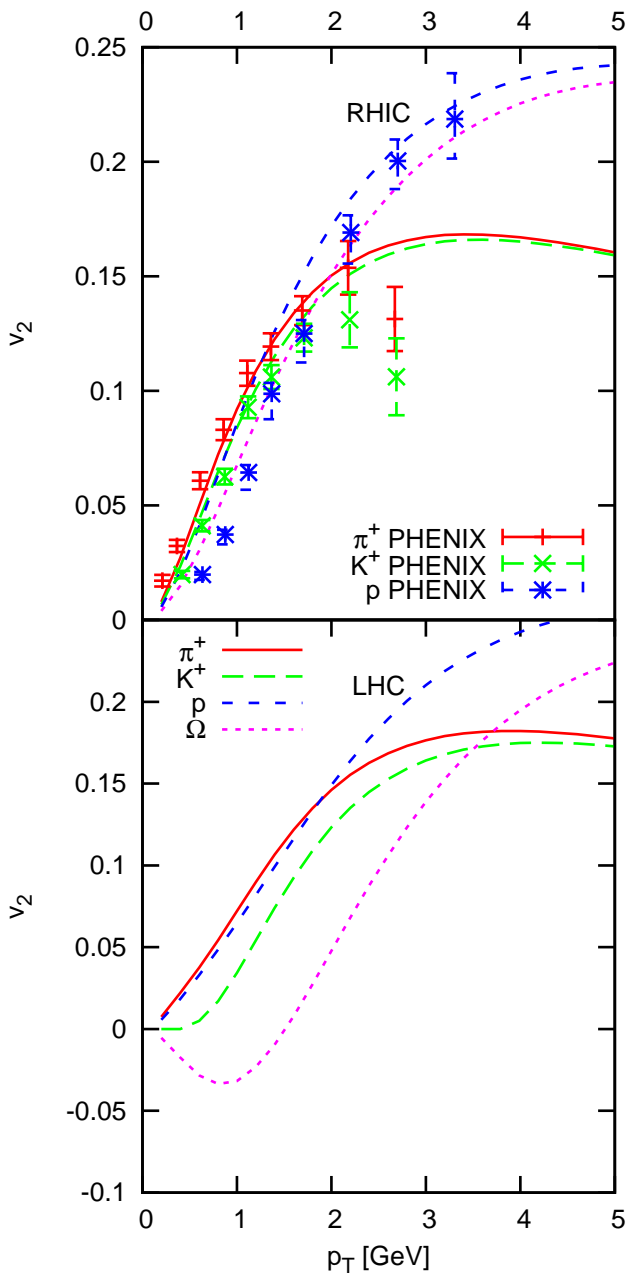}
 \caption{(Color online) Elliptic flow parameter $v_2$ as a function of the transverse momentum $p_t$ for mid-central ($b=8\fm$) 
Au+Au/Pb+Pb  reactions as obtained from the recombination approach. The lines indicate the calculations for 
for $\pi^+, K^+, p \mbox{ and } \Omega^-$. 
\textbf{Upper Panel:} $v_2$ at RHIC ($\sqrt{s}=200 \GeV$) compared to PHENIX data \cite{Adler:2003kt}. 
\textbf{Lower Panel:} Prediction for LHC ($\sqrt{s}=5.5 \TeV$).}
 \label{plt:pt-v2}
\end{figure}
with
\[a(\phi) = p_T \sinh(\eta_T(\phi)) /T,\quad b(\phi) = m_T \cosh(\eta_T(\phi)) /T\]
and the modified Bessel functions $I_n$ and $K_n$. Replacing $w_a(R;p)$ in Eq.(\ref{eqn:main}) with
\begin{equation}
w_a'(R;p) = w_a(R;p) \left(1+2v_2(p_T) \cos (2\Phi)\right)
\end{equation}
and again applying the definition for $v_2$ yields the hadronic elliptic flow.

Let us start by comparing the transverse momentum dependence of the elliptic flow for pions, kaons, protons 
and omega-baryons in mid-peripheral Au+Au reactions at RHIC to the distributions obtained in mid-peripheral
Pb+Pb reactions at the LHC as shown in Fig. \ref{plt:pt-v2}. 
A first striking observation is that the $v_2(p_T)$ values at each given $p_T$ are always lower at the LHC 
than at RHIC.  A similar pattern was also observed within a parton transport approach, if the viscosity 
was set to the ADS/CFT limit  \cite{Molnar:2007an}. Even more surprising, we find strong out-of-plane emission
of (multi-)strange particles, i.e. negative values of $v_2$. 
These negative values of the elliptic flow parameter for heavy particles have also been found 
in previous exploratory studies and seem to be a general feature of the blast-wave like flow profile
at high transverse velocities \cite{Voloshin:1996nv,Huovinen:2001cy,Retiere:2003kf,Pratt:2004zq}. 
The negative $v_2$ values have also be seen within a blast-wave fit of the proton 
elliptic flow to STAR data \cite{Voloshin:2002ii} and they have also been experimentally observed 
by the NA49 collaboration \cite{Alt:2003ab}. These results are supported by the recent 
preliminary data on J/$\psi$ elliptic flow from PHENIX \cite{Krieg:2008bc}.
It reflects the depletion of the low $p_T$ particle abundance, when the source elements are highly boosted
in the transverse direction. At higher $p_T$ and or lower transverse flow velocities, the opposite effect 
dominates and results in a more pronounced emission in-plane again. 
The details  depend on the underlying transverse flow parameterisation as discussed in \cite{Pratt:2004zq}.
One might argue that negative elliptic flow values are an artefact of the blast wave parametrisation, however, also
transport simulations indicate slightly negative $v_2$ values for heavy particles at 
very low $p_T$ \cite{Bleicher:2000sx}. Thus, the qualitative behaviour
 of a negative $v_2$ is a well-known observation. The surprising observation in the
present work is the quantitative strength of this effect at LHC.

From these arguments we conclude that the blast-wave flow profile is responsible for the 
negative $v_2$ on the quark level, which then enters 
the hadron elliptic flow via the parton recombination. To understand this effect in more detail we go to the thermal 
quark-spectrum \cite{Fries:2003kq} with the energy
\begin{eqnarray}E(R,p) &=& m_T \cosh(\eta-y) \cosh \eta_T(\phi) \nonumber\\*
&&- p_T \cos(\phi-\Phi) \sinh \eta_T(\phi)
\end{eqnarray}
For simplicity we look at midrapidity ($y=0$) and for high $\eta$ the spectrum is very low, so we consider the region around $\eta = 0$. For high cm-energies, when the source is highly boosted transversally, the particles will mainly be emitted in the direction in which the fireball flies, so we can simplify even more and set $\phi = \Phi$. Because we have no longitudinal momentum we can replace the momentum with the transverse rapidity of the parton: $p_T=m \sinh \eta_T^q$
and $m_T = \sqrt{m^2+p_T^2}= m\sqrt{1+\sinh^2 \eta_T^q} = m \cosh \eta_t^q$. So the energy of the quark in the transverse moving source is
\begin{equation}
E = m \cosh\left[\eta_T^0(1+ \alpha f(p_T) \cos(2\phi))-\eta_T^q \right]
\end{equation}
For a fixed quark rapidity $\eta_T^q < \eta_T^0$ (low $p_T$) the energy of the quarks emitted in plane is higher than the energy of quarks emitted out of plane. With the thermal spectrum more energy means less particles, therefore a negative $v_2$. At $\eta_T^q = \eta_T^0$ one would expect the zero-crossing, and above a positive $v_2$.

In this way one can visualise our predicted negative elliptic flow for low $p_T$ at LHC. At RHIC this picture can not be applied, because the simplification $\phi = \Phi$ is only valid for high $\eta_T^0$ (high cm-energies).

A different way to obtain an analytic expression which explains the negative $v_2$ is to consider only the 
in-plane ($\phi=0$) and out-of-plane ($\phi=\pi/2$) directions (similar to the analysis performed 
in \cite{Huovinen:2001cy}). The $\phi$ 
integration breaks down and the elliptic flow is then given by
\begin{equation}
v_2^q(p_T) = \frac{I_2 \left[a(0)\right] K_1 \left[b(0)\right] - I_2 \left[a(\pi/2)\right] K_1 \left[b(\pi/2)\right]}
{I_0 \left[a(0)\right] K_1 \left[b(0)\right] + I_0 \left[a(\pi/2)\right] K_1 \left[b(\pi/2)\right]}
\end{equation}
For $p_T \rightarrow 0$ the argument of the Bessel functions $I_n$ goes to zero and $I_n$ becomes 
constant with $I_2 \rightarrow 0$ and $I_0 \rightarrow 1$. Therefore they are independent of the angle. This leads to
\begin{eqnarray}
&\lim_{p_T \rightarrow 0} v_2^q(p_T) = \frac{I_2 \left[a(0)\right]}{I_0 \left[a(0)\right]}\nonumber\\
\times &\frac{K_1 \left[m \cosh(\eta_T^0(1+\alpha)))/T\right] - K_1 \left[m \cosh(\eta_T^0(1-\alpha)))/T\right]}
{K_1 \left[m \cosh(\eta_T^0(1+\alpha)))/T\right] + K_1 \left[m \cosh(\eta_T^0(1-\alpha)))/T\right]}
\end{eqnarray}
Since $K_1$ is a monotonically decreasing function the nominator the elliptic flow is negative (for some small 
transverse momenta). The specific values depend on the mass, the mean flow rapidity $\eta_T^0$ and the temperature. 
For increasing mass, or increasing transverse flow rapidity or decreasing temperature, the elliptic flow 
will become more negative.

\begin{figure}
 \centering
 \includegraphics{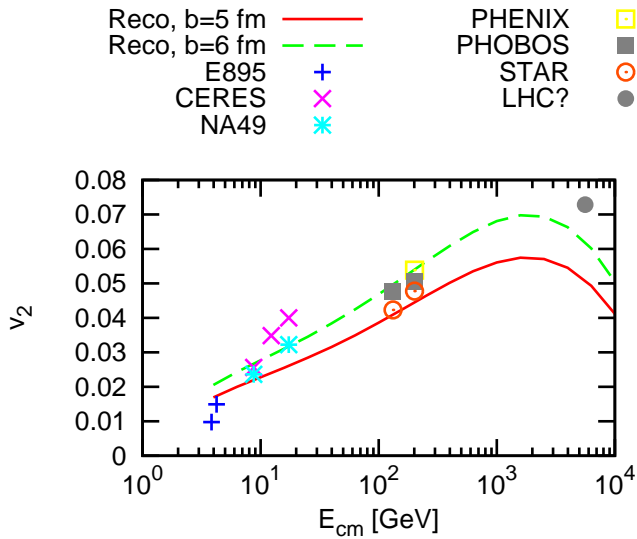}
 \caption{(Color online) Comparison of $\left\langle v_2 \right\rangle$ for charged hadrons at midrapidity as a function of center of mass energy from the present
recombination approach for Au+Au/Pb+Pb reactions at an impact parameter of $b=5-6\fm$. Data points and the extrapolation to LHC are taken from \cite{Borghini:2007ub}.}
 \label{plt:v2_E}
\end{figure}

Folding the distributions for $v_2(p_T)$ with the corresponding transverse momentum distributions 
yields the integrated $v_2$ at midrapidity for each particle species, see Fig. (\ref{plt:v2_E}). 
A first observation is the apparent good description of the available charged particle $v_2$ data from the AGS to the
RHIC energy regime.  
However, when going above the highest RHIC energy, we predict that the integrated elliptic flow saturates and
then starts to {\it decrease} in the LHC energy range. This finding is in stark contrast to the current expectations
in the heavy ion community - compare e.g. to the linear extrapolations towards LHC by Borghini and 
Wiedemann \cite{Borghini:2007ub}). 

\begin{figure}
 \centering
 \includegraphics{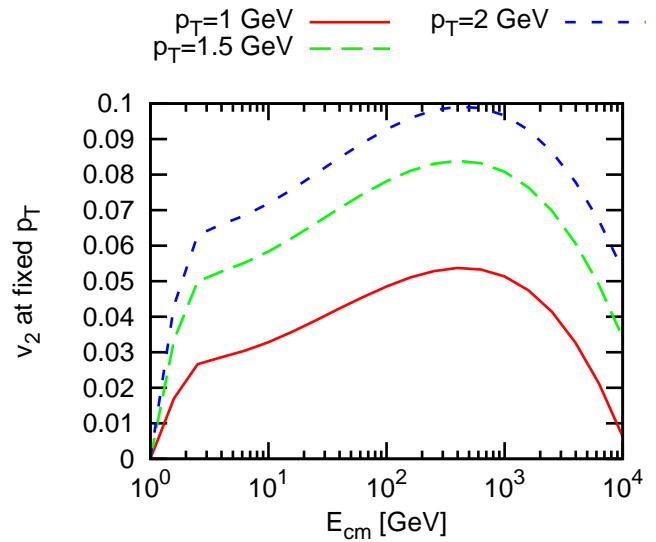}
 \caption{(Color online) Elliptic flow $v_2$ of charged hadrons at $b=5-6$ and fixed $p_T$ as a 
function of $\sqrt{s_{NN}}$ for $p_T = 1, 1.5 \mbox{ and } 2 \GeV$.}
 \label{plt:v2_fixed_pt}
\end{figure}

A rather critical assumption in this study is the applicability of the recombination approach 
for the elliptic flow for small transverse momenta on the order of $p_T<1 \GeV$. We want to 
emphasise that our result of the decreasing mean 
elliptic flow  $\langle v_2\rangle$ at LHC is not affected by the validity of this assumption, 
because $\langle v_2\rangle \approx  v_2\left(\langle p_T\rangle\right)$ and $\langle p_T\rangle > 1 \GeV$ in LHC regime. 
However, to show the robustness of the prediction Fig. (\ref{plt:v2_fixed_pt}) depicts the elliptic flow 
at a fixed $p_T$ as a function of $\sqrt{s}$. With $p_T=1, 1.5 \mbox{ and } 2\GeV$ the elliptic flow exhibits the same 
drop as the mean $v_2$ from Fig. (\ref{plt:v2_E}).

We have calculated the energy dependence and transverse momentum dependence of 
the elliptic flow parameter $v_2$ for mid-central Au+Au/Pb+Pb collisions from AGS to LHC energies 
within a parton recombination approach. We find a reasonable description of the $v_2$ data over the 
whole inspected energy regime, indicating that the measured $v_2$ values are consistent with the assumption 
that a major part of the elliptic flow was created in the partonic stage.
We predict that the integrated $v_2$ of charged particles at midrapidity will {\it decrease} from RHIC to LHC energies,
due to the strong transverse flow.
In detail, we link this to the prediction of a negative $v_2$ component developing at low transverse momenta to 
the blast-wave like flow profile of the underlying quark distribution. 
Because this effect is strongest for heavy quarks it most visible  for (multi-)strange hadrons. 
Above the presently envisaged LHC energy for nucleus-nucleus reactions,
we predict that the mean elliptic flow will even turn negative. It should be pointed out that the present prediction is
in striking contrast to all former assumptions about the behaviour of $v_2$ at LHC.

\section*{Acknowledgements}
This work was supported by Gesellschaft f\"ur Schwerionenforschung, Darmstadt (GSI). The Authors would like 
to thank Drs. Nu Xu and Steffen A. Bass for fruitful discussions.

\appendix
\section*{Appendix A}

\begin{figure}
 \centering
 \includegraphics{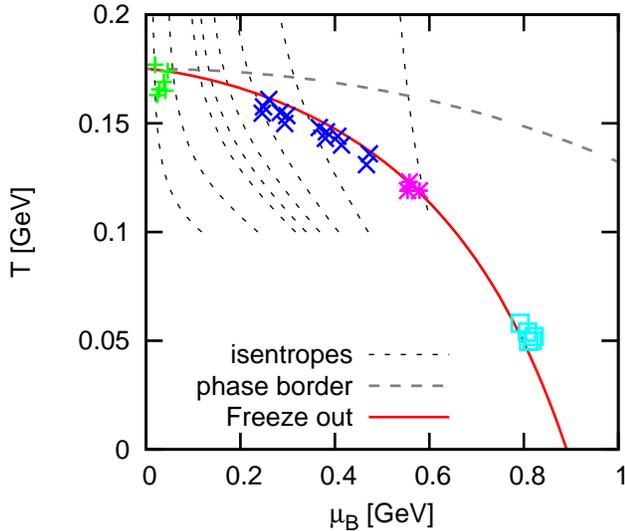}
\caption{(Color online) The QCD phase diagram with a freeze-out curve fitting the collider experiments (RHIC, SPS, SIS, AGS) and some isentropes to the MIT-Bad model phase boundary.}
 \label{plt:phase_diagram}
\end{figure}

\begin{figure}
 \centering
 \includegraphics{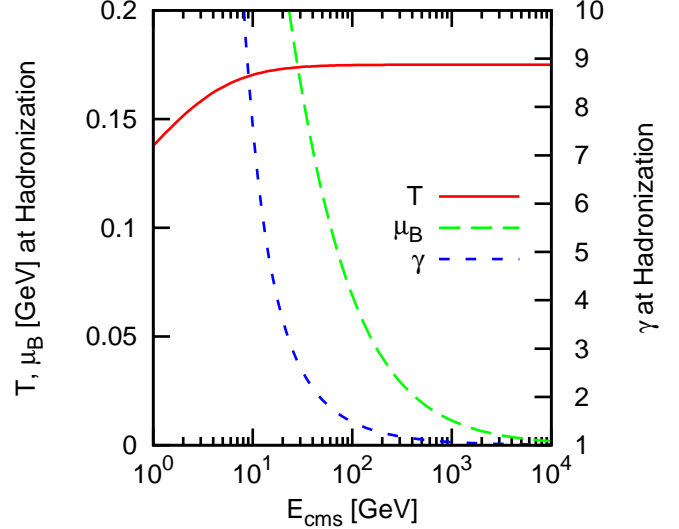}
\caption{(Color online) Temperature $T$ and baryo-chemical potential $\mu_B$ (left y-axis) and fugacity $\gamma=e^{(\mu_B / T)}$ 
(right y-axis) at the phase transition as a function of the center of mass energy $\sqrt{s_{\rm NN}}$.}
 \label{plt:Tmu}
\end{figure}

To extract $T$ and $\mu_B$ at the phase boundary we take the freeze-out values of the baryo-chemical potential from different accelerators \cite{Cleymans:2006zz} and fit them against their cm-energy via
\begin{equation}\mu_B^{\rm Freeze}=\frac{a}{1+\sqrt{s_{\rm NN}}^b/c}\end{equation}
with $a = 1.478 \GeV,\; b = 0.802, \; c = 2.096 \GeV$. Then we follow the isentropes from freeze-out to the phase boundary in order to connect this $\mu_B^{\rm Freeze}$ to a $\mu_B$ value at the phase boundary (Fig. (\ref{plt:phase_diagram})). In the relevant region, the connection is approximately linear with
\begin{equation}\mu_B = d \cdot \mu_B^{\rm Freeze}, \;d = 0.938\end{equation}
Due to the lack of lattice data at high baryo-chemical potentials, we calculate the phase transition line with a simple MIT-Bag model with a critical temperature of $T_C = 175 \MeV$ at vanishing chemical potential. The line of the phase transitions temperature is
\begin{equation}
T =\sqrt{-\left(\frac{\mu_B}{3}\right)^2 \frac{g_q}{C}+\sqrt{\left(\frac{\mu_B}{3}\right)^4 \frac{g_q}{C^2}\left(g_q-\frac{C}{\pi^2}\right)+T_C^4}}\\
\end{equation}
with $C=\frac{\pi^2}{15}(7g_q+4g_g), \; g_q = 12,\; g_g = 16$. Details like the order and nature of the parton-hadron transition are not relevant for the 
present study. Because the fugacity $\gamma = \exp(\mu_B /T)$ does not 
enter in the equation for $v_2$ (in any case $\gamma \simeq 1$ for all relevant flavours at LHC energies), we neglect 
it in this paper. The temperature, baryo-chemical potential and the fugacity at hadronization are 
shown in Fig. (\ref{plt:Tmu}).

\section*{Appendix B}

\begin{figure}
 \centering
 \includegraphics{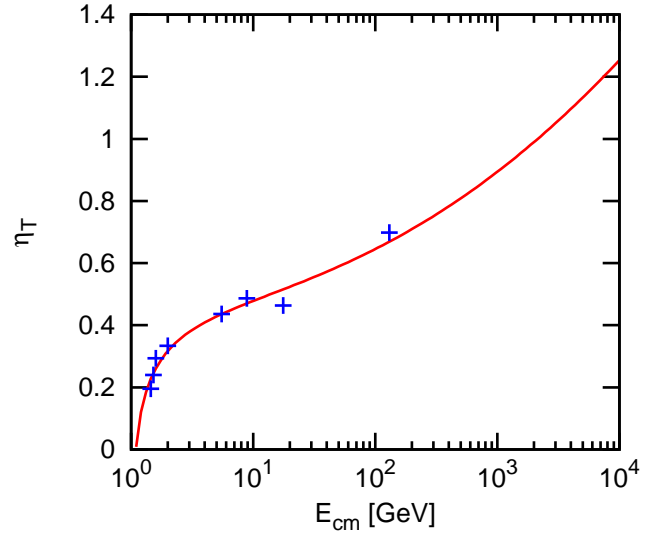}
 \caption{(Color online) Parameterisation of $\eta_T$ (full line) as function of $\sqrt{s}$. 
The data at kinetic freeze-out (crosses) are taken from \cite{Xu:2001zj}.}
 \label{plt:eta_T}
\end{figure}

While the hadronization temperature shows only a weak dependence on energy, the transverse expansion rapidity $\eta_T$ 
is strongly energy dependent. We fit the flow rapidities extracted from the experimental data \cite{Xu:2001zj} 
at kinetic freeze-out. The data given in \cite{Xu:2001zj} is for mean transverse flow velocity $\beta_T$, 
but we use the rapidity $\eta_T=\tanh \beta_T$ to assure that the velocity stays less than $c$. 
To fit these values we choose 
\begin{equation}\eta_T^{\rm Freeze}(\sqrt{s}) = a + b x + c x^2 + d \ln(x)\quad, x=ln\left(\sqrt{s}\right)\end{equation}
with the constants $a = 0.418, \; b = -0.064,\; c=0.012,\; d=0.170$.
As these values are extracted at freeze-out, we scale the obtained transverse rapidities by a constant 
factor $k=0.85$ to obtain the transverse flow at the hadronization surface.
Using these parameters our value for $v_T=0.54$ at RHIC energies agrees with the value from \cite{Fries:2003kq}. 
For the LHC energy ($\sqrt s =5.5$~TeV) we obtain a transverse flow velocity of $v_T=0.75$ also in line with previous 
estimates \cite{Fries:2003fr}. Fig. (\ref{plt:eta_T}) depicts our fit (line) and the available 
data on $\eta_T$ (crosses).

\end{document}